\documentclass[journal=jpcafh,manuscript=article]{achemso}

\usepackage[version=3]{mhchem} 
\usepackage[dvipsnames]{xcolor}
\usepackage{bm}
\usepackage{threeparttable}
\usepackage{amsmath,amsfonts,amssymb}
\usepackage{booktabs}
\usepackage{graphicx,caption}
\usepackage{soul}

\newcommand{\cm}{cm$^{-1}$ }
\author{Paul L. Houston}
\affiliation{Department of Chemistry and Chemical Biology, Cornell University, Ithaca, New York
14853, U.S.A. and Department of Chemistry and Biochemistry, Georgia Institute of
Technology, Atlanta, Georgia 30332, U.S.A}
\email{plh2@cornell.edu}
\author{Chen Qu}
\email{szquchen@gmail.com}
\affiliation{Independent Researcher, Toronto, CA}

\author{Qi Yu}
\email{qyu28@emory.edu}
\affiliation{Department of Chemistry and Cherry L. Emerson Center for Scientific Computation, Emory University, Atlanta, Georgia 30322, U.S.A.}
\author{Priyanka Pandey}
\affiliation{Department of Chemistry and Cherry L. Emerson Center for Scientific Computation, Emory University, Atlanta, Georgia 30322, U.S.A.}

\author{Riccardo Conte}
\affiliation{Dipartimento di Chimica, Universit\`{a} degli Studi di Milano, via Golgi 19, 20133 Milano, Italy}

\author{Apurba Nandi}
\affiliation{Department of Chemistry and Cherry L. Emerson Center for Scientific Computation, Emory University, Atlanta, Georgia 30322, U.S.A.}
\alsoaffiliation{Department of Physics and Materials Science, University of Luxembourg, L-1511, Luxembourg City, Luxembourg}

\author{Joel M. Bowman}
\email{jmbowma@emory.edu}
\affiliation{Department of Chemistry and Cherry L. Emerson Center for Scientific Computation, Emory University, Atlanta, Georgia 30322, U.S.A.}

\title{\textit{No Headache for PIPs: A PIP Potential for Aspirin Outperforms Other Machine-Learned Potentials}}

\begin{document}

\begin{abstract}
Assessments of machine-learned (ML) potentials are an important aspect of the rapid development of this field.  We recently reported an assessment of the linear-regression permutationally invariant polynomial (PIP) method for ethanol, using the widely-used (revised) MD17 dataset. We demonstrated that the PIP approach outperformed numerous other methods, e.g., ANI, PhysNet, sGDML, p-KRR, with respect to precision and notably with respect to speed [Houston $et$ $al$., \textit{J. Chem. Phys.} \textbf{2022}, \textit{156}, 044120.].  Here we extend this assessment to the 21-atom aspirin molecule, using the rMD17 dataset. Both energies and forces are used for training and the precision of several PIPs is examined for both.  Normal mode frequencies, the methyl torsional potential and 1d vibrational energies for an OH stretch are presented. Overall,  we show that the PIPs approach outperforms other ML methods, including sGDML, ANI, GAP, PhysNet and ACE, as reported by Kov{\'a}cs $et$ $al.$ in \textit{J. Chem. Theory Comput.} \textbf{2021}, \textit{17}, 7696–7711. 
\end{abstract}   

\section{Introduction}
Permutationally invariant polynomials (PIPs)\cite{Braams09, Xie10,Houston2023PIPsoftware} have been used for nearly 20 years to develop precise, high-dimensional potential energy surfaces (PESs) for molecules.\cite{Bowman2011,ARPC2018,czako2021}  The precision of a PIPs PES for rMD17 ethanol was shown to be as good as the best performing ML methods and to be substantially faster (factors of 10 or more) than all the ML methods considered, i.e., GAP-SOAP,\cite{GP-2015-1} ANI,\cite{AN1} DPMD,\cite{dpmd2018} sGDML,\cite{Tkatch2018, Tkatch19} PhysNet,\cite{PhysNet} KREG,\cite{KREG} and pKREG \cite{pKREG}.  

It is of course important to continue assessing  ML methods in terms of precision and speed for ever-increasingly large molecules.  Kov{\'a}cs et al recently presented such an assessment for a number of molecules in the rMD17 dataset.  While all methods performed reasonably well, they all produced the largest precision errors for energies and forces for 21-atom aspirin.\cite{Kovacs2021}  This is a major motivation for us to select aspirin among the few 20+-atom molecules in the rMD17 dataset; another is the challenge of describing the torsional potential of the methyl rotor.  Thus, we report several PIP PESs for aspirin using the rMD17 dataset and show the most precise one outperforms all the ML methods considered by Kov{\'a}cs \textit{\textit{et al}.}, including ACE, the method developed  by Kov{\'a}cs \textit{et al}., in a manner very similar to the performance for rMD17 ethanol mentioned above.  This is especially important for PIPs which has been claimed erroneously to be limited to ``small molecules" in the literature.  

Acetylsalicylic acid (ASA), known as Aspirin, is an anti-inflammatory drug used to treat pain or fever.  Derived originally from plants such as the willow, it has been used since as early as 1550 BCE.  It's mechanism was explained by J. R. Vane in 1971\cite{VAne1971} as being due to its ability to inhibit prostaglandin synthesis, a discovery for which he shared the Nobel Prize in Physiology or Medicine in 1982.  Quantum Mechanics/Molecular Modelling calculations have recently produced a reduced dimensional potential energy surface (PES) to study how the ASA acetylates a serine site on the COX-1 enzyme, the reaction known to effect the inhibition.\cite{Toth2013} 

The potential energy surface of aspirin itself has recently been important in the assessment of machine learned (ML), full-dimensional potential energy surfaces, where it has become a benchmark.\cite{Chmiela2017,Chmiela2018,Tkatchenko2021,Kovacs2021,DralPinheiro2021,DralHou2023} With 21 atoms, it is close to the largest molecule in the MD17 and rMD17 database.\cite{Chmiela2017,Christensen2020}  In the context of our own work, aspirin is larger than the largest molecules for which we constructed and reported PIP PESs, namely 15-atom acetylacetone \cite{QuAcAc2020,QuAcAc2021,Nandi2023JACS} and tropolone.\cite{Nandi2023JACS}  

Perhaps more importantly, aspirin provides a test case for evaluating some of the most accurate and cost-effective ML methods for constructing potential energy surfaces.  A recent article\cite{Kovacs2021} has shown that ACE and PaiNN\cite{Schutt2021} are the most accurate ML methods of those considered.   Most recently, the atom-centered message passing neural network ``Allegro'' method has achieved the smallest MAE for energies and forces.\cite{Musaelian2023} as we will show below, our own method based on PIPs outperforms ACE and PaiNN and and is essentially the same as Allegro in MAE.  We show explicitly that the PIPs potential is roughly 40 times faster than the ACE potential which is faster than PaiNN.  We don't have timings of Allegro for the Asprin fit, however, this method almost certainly runs slower than ACE and so much slower than the PIPs potential.


As discussed in the first papers on fragmentation,\cite{QuBowman2019,NandiQuBowman2019} Morse variables decrease exponentially as the corresponding internuclear distances increase. Thus, there are polynomials, especially for large molecules, that are small and can be discarded, so that a very large basis may be greatly reduced in size by discarding PIPs that are uniformly very small over the dataset.  We term this a ``pruned basis". Of course this requires the calculation of the large PIPs basis, which in some instances for large molecules may require a very large cpu effort.  So instead of doing that, a fragmented basis was proposed.  We don't do that here because it is possible to generate two PIPs bases for aspirin. The much larger basis was pruned then using methods described in elsewhere,\cite{Houston2023PIPsoftware} and in detail below. The smaller basis is too small given the size of the dataset and so we use a simple method to increase the size of that basis.  This is also described in detail below.


The remainder of the paper consists of a section on the dataset; the generation of the PIPs bases; followed by results, which include the usual metrics; and comparisons with other ML PESs with respect to precision and timing. Then a number of properties of the new PIP PES are presented starting with the methyl torsional barrier and potential.  Behavior of  the PES beyond the range of the data is examined for a local mode O-H stretch (including 1d quantum vibrational calculations of that mode) and two internal rotational motions.  A short discussion contrasting the PIPs approach with atom-center ones is then given, followed by a summary and conclusions. 

\section{Methods}
\subsection{Dataset}
\begin{figure}[htbp!]
\begin{center}
\includegraphics[width=.6\textwidth]{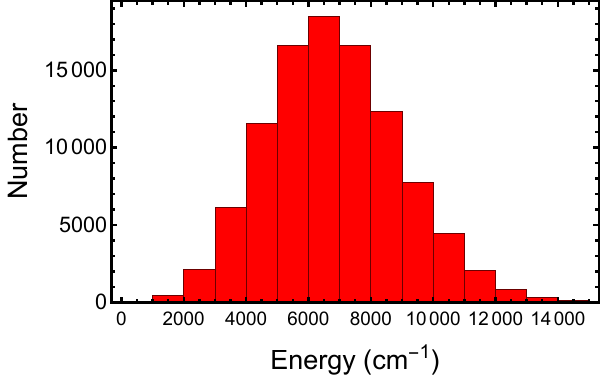}
\end{center}
\caption{Histogram of aspirin energies from the rMD17 dataset}
\label{fig:histogram}
\end{figure}

Our calculations, like those to which we will compare them, makes use of the rMD17 database.\cite{Chmiela2017,Christensen2020} This set of electronic structure calculations was performed with ``direct-dynamics" using Density Functional Theory with the PDE functional  and the def2-SVP basis.  The total number of geometries included for aspirin is 100,000, and both the energy and the gradient components were calculated for each geometry.  The energy distribution for these geometries is given in the histogram of Fig. \ref{fig:histogram}.  This extends to almost 14 000 cm$^{-1}$.  This distribution is just the expected, thermal distribution of potential energies of 57 (i.e., $3 \times 21-6$) harmonic oscillators at approximately 350 K, with an average value of 5933 cm$^{-1}$ in rough agreement with the actual average in the figure of 6827 cm$^{-1}$. Thus, it is reasonable to characterize this as the potential energy partitioned among 57 classical oscillators, with an average of roughly 100 \cm and a maximum of roughly 245 \cm per oscillator.

The primary purpose for the rMD17 database was to aid in evaluating machine learned potentials, so the chemical deficiencies of the database may perhaps be overlooked.  These limitations have been examined in detail by us in a recent Perspective article,\cite{Bowman2022} and have mostly to do with the range of the energies included. Because the zero-point energy of aspirin is much higher than the upper limit of the data (the harmonic zero-point is 33,500 cm$^{-1}$), it is clear that the rMD17 energy range would be inadequate for calculating the actual zero point energy by, say, diffusion Monte Carlo techniques.  At the lower end of the energy range, we note that there are only 46 geometries with energies below 1000 cm$^{-1}$.  This makes it unlikely that the lower vibrational frequencies will be accurately determined, or that there would be enough points to describe low frequency motions, such as the hindered rotation of the methyl group addressed below.  The main advantage of the dataset is that there are a large number of data points to fit, 100,000 energies and 6,300,000 force or gradient components. Although there is significant correlation/redundancy in this data set, this is not an issue for our fitting method, which is overdetermined least squares regression in terms of  permutationally invariant polynomials.  In one equation this is given by
\begin{equation}
V(\textbf{y})= \sum_{i=1}^{n_p} c_i p_i(\textbf{y}),
\label{eq1}
\end{equation}
where $c_i$ are linear coefficients, $p_i$ are PIPs, $n_p$ is the total number of polynomials (and linear coefficients $c_i$) for a given maximum polynomial order, and $\textbf{y}$ are transformed internuclear distances $r_{ij}$ between atoms $i$ and $j$, i.e., $y_{ij}$=$exp(-r_{ij}/a)$.\cite{Braams09}. The range parameter $a$ is typically between 2 and 3 bohr.

\subsection{Permutationally Invariant Polynomial Bases}
\begin{figure}[htbp!]
\begin{center}
\includegraphics[width=.6\textwidth]{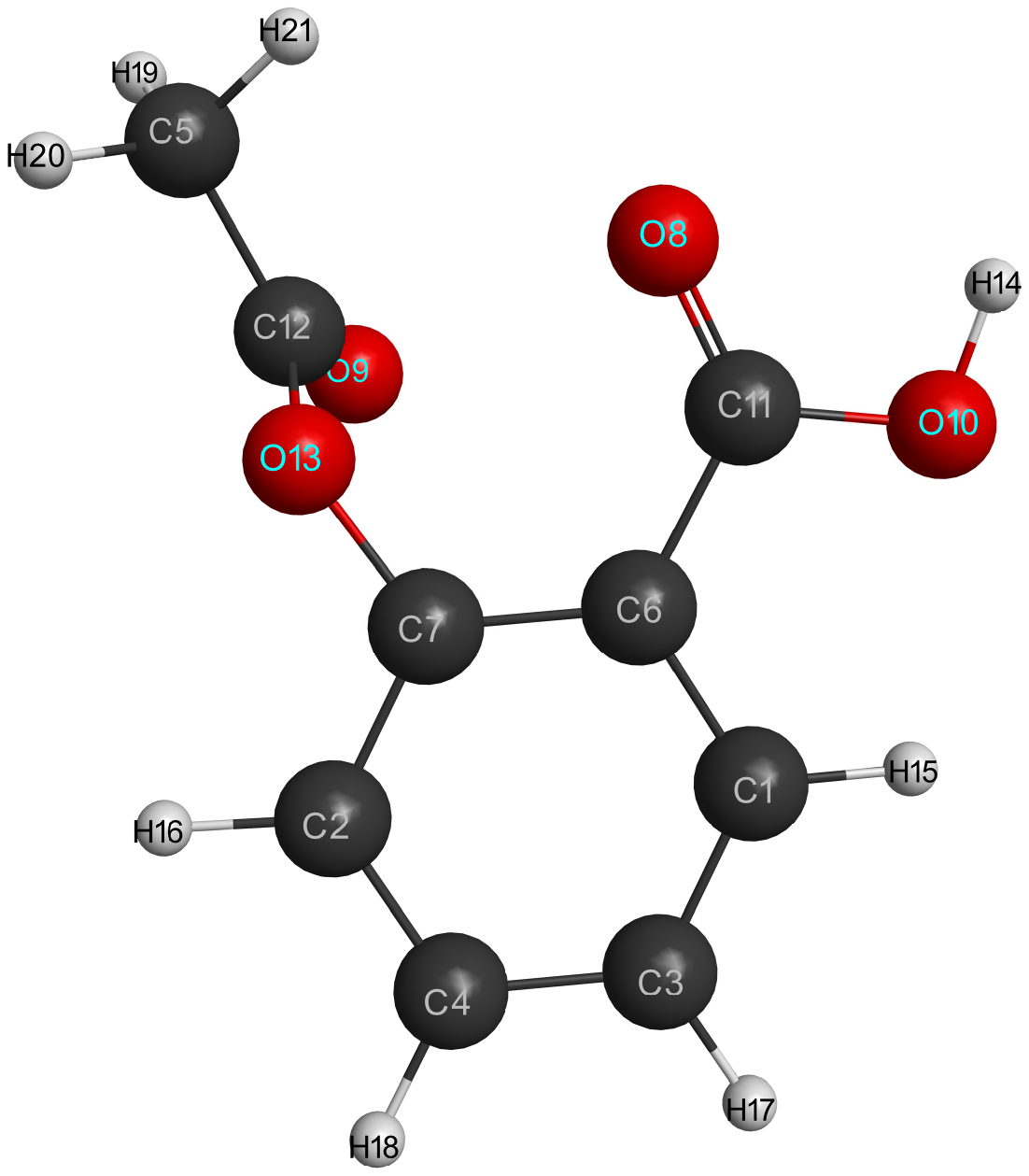}
\end{center}
\caption{Ball and stick model of aspirin, showing the atom names corresponding to the rMD17 dataset positions in Table \ref{tab:correspondence}}
\label{fig:ballstick}
\end{figure}

Figure \ref{fig:ballstick} shows a ball and stick diagram of aspirin.  The molecule has an acetyl group attached at C7 of the benzene ring and a formic acid group attached at C6.  In the equilibrium structure shown, the benzene and the formic acid are co-planar, whereas the acetyl group is twisted out of the plane. The numbering shown is related to the individual members of the dataset as well as to the permutational symmetry designation for the molecule, where the correspondence is shown in Table \ref{tab:correspondence}.

\begin{table}[htbp!]
    \centering
    \begin{tabular}{l c c c}
    \toprule
    Atom & dataset & Symmetry & Symmetry \\
         & Position & Position & Group \\
    \midrule
     H6 & 19 & 1 & 3 \\
     H7 & 20 & 2 & 3 \\
     \vspace{2mm}
     H8 & 21 & 3 & 3 \\
     O1 & 8 & 4 & 2 \\
     \vspace{2mm}
     O3 & 10 & 5 & 2\\
     C3 & 3 & 6 & 2\\
     \vspace{2mm}
     C4 & 4 & 7 & 2\\
     C1 & 1 & 8 & 2\\
     \vspace{2mm}
     C2 & 2 & 9 & 2\\
     H2 & 15 & 10 & 2\\
     \vspace{2mm}
     H3 & 16 & 11 & 2\\
     H4 & 17 & 12 & 1\\
     H5 & 18 & 13 & 1\\
     C5 & 5 & 14 & 1\\
     C6 & 6 & 15 & 1\\
     C7 & 7 & 16 & 1\\
     C8 & 11 & 17 & 1\\
     C9 & 12 & 18 & 1\\
     O2 & 9 & 19 & 1\\
     O4 & 13 & 20 & 1\\
     H1  & 14 & 21 & 1\\
    \bottomrule
    \end{tabular}
    \caption{Correspondence of Atom Positions}
    \label{tab:correspondence}
\end{table}

Our method for determining the PES uses a set of PIPs whose values at any geometry form a vector.  The energy is then calculated by the vector product of the polynomial values and a vector of coefficients to be determined so that the energies  fit those of the data set.  The vector product of the same coefficients with the derivatives of the polynomials is also required to give the components of the gradients or forces.  The permutational invariance of the polynomials ensures invariance of the potential with respect to permutation of like atoms. Of course restrictions can be made on whether a permutation is energetically feasible or not, and this is user-determined when calling our software to generate the polynomials.  For example, the three hydrogen atoms on the methyl group are treated as identical  in order to correctly describe the torsional potential; it must have three-fold symmetry.  While they should also, in principle, commute with other hydrogens in the molecule, it is unlikely at the energies of this study that such processes would occur at a rate sufficient to be observed, so not all permutations need to be included. At the same time, using permutational symmetry will reduce the number of unique polynomials and lead to a more compact description of the PES, one that can be evaluated rapidly.  There is trade-off between using maximum permutational symmetry, which results in the smallest number of basis functions but which can be computationally expensive to evaluate, and using lower but essential symmetry, resulting in more functions but which are less intensive to evaluate. In addition, the cpu effort to generate PIPs with maximum permutational symmetry can be considerably larger than than for a basis with lower symmetry.

One important aspect of our PIP software ``MSA" is that it starts with a user-defined symmetry.\cite{msachen,Xie-nma} The output of this software has the added advantage that calculating the polynomial values given a geometry is designed to be computationally efficient. For aspirin the symmetry 322221111111111 was selected. This means that the basis functions are invariant with respect to permutations of three atoms (the methyl H atoms), as well as four groups of two atoms, and that the remaining 10 atoms that are treated as distinguishable. These groups are shown by the vertical separations in Table \ref{tab:correspondence}. The first group has three members, hydrogens 19, 20, and 21 in the figure.  These are the last three atoms listed in the geometry descriptions of the data set and the first three listed in the permutational symmetry. There are then four groups of two identical atoms that : O1 and O3, C3 and C4, C1 and C2, and H2 and H3.  Each atom has a position in the dataset and a position in the symmetry description.  The final ten atoms are listed at the bottom of the table; these do not permute with any others in this description of the symmetry.

Using this symmetry with a maximum polynomial order of 2 and 3 results in bases of 6766 polynomials/coefficients and 297,952 polynomials/coefficients.  The second order basis is small considering the size of the dataset, so it was increased as described briefly below. The third order basis is too large for our computational resources (memory).  Also, based on our experience the number of coefficients/polynomials should be 3 or more times smaller than the number of energy and gradient constraints so as to avoid over-fitting, a situation in which all the points are well fit but the region between them has strong oscillations in energy. So this basis is pruned using recent PESPIP software\cite{Houston2022PIPs,Houston2023PIPsoftware} (available at 
https://github.com/PaulLHouston/PESPIP). The software offers a solution to these problems by generating PIP basis sets that either add or prune polynomials to/from an existing basis. In our work, the polynomials are usually functions of Morse variables, although reciprocal internuclear distances are also an option.  The reason for these is that they allow the geometrical distances to be denoted by a number between 0 and 1, and thus put different distances on the same numerical footing. The Morse and inverse variable functions are, respectively, $\exp(-r_{ij}/a)$ or $1/r_{ij}$, where $r_{ij}$ is the internuclear distance between atoms $i$ and $j$ and $a$ is a range parameter generally taken to be about 2 bohr. The PESPIP software generates new polynomials by multiplying together combinations of the existing PIPs.  We note that the product of two PIPs is still a polynomial that is permutationally invariant. The software prunes polynomials from an existing set by deleting the least important polynomials and reorganizing the result so that the remaining ones are still permutationally invariant and still may be calculated efficiently.  It remains to describe how the choice of which polynomials to add or prune is made.

The criterion for keeping a polynomial is based on the fact that the most important polynomials are those with the largest values when evaluated using the geometries of the data set.  (Other methods, such as fragmentation,\cite{QuBowman2019} might alternatively be used in a manner to maintain the desired permutational symmetry.) The procedure is to evaluate the maximum values of the newly calculated polynomials, in the case of addition, or the existing ones, in the case of pruning, for each geometry in the data set, keeping the maximum value associated with each polynomial. We next make a list of polynomials ordered from maximum to minimum value.  Polynomials associated with small internuclear distances have large values and will be at the top of the list, while those associated with large internuclear distances have small values and will be at the bottom.  By keeping those polynomials with large values, we focus on atoms that are close to one another and partially ignore those that are far away from one another. For polynomial addition, we generate the list as we multiply existing polynomials, always being sure that we consider multiplication of the largest-value polynomials first and making sure that the generated polynomial is not already on the list.  When we reach the desired number of polynomials we stop.  For pruning of polynomials, we evaluate the existing polynomials in an ordered list and go down the list until we we reach to the desired number of polynomials; we prune the rest. More details may be found in reference \citenum{Houston2023PIPsoftware}.  We will examine the resulting PESs using these methods in the Results section. 

In the end we used the PESPIP software to increase the $2^{nd}$-order set of PIPS from 6766 to 25 001 and to prune the $3^{rd}$-order set of PIPS from 297 952 to 49 978. These have small numbers of coefficients compared to the full MD17 dataset which consists of 100 000 energies and 630 000 gradient components. The PESPIP software took about 53 minutes to augment the $2^{nd}$-order set and about 170 minutes to prune the $3^{rd}$-order set. Energies were weighted by E/(E+$\Delta$), where $\Delta$ = 0.02 hartree, and the gradients are additionally weighted by a factor of 1/3.  More comments on these aspects of the PIP fits are given below.


\section{Results}
We generally do not use the train/test protocol, where a given dataset is split randomly into a training and testing set. (We did this in a 2019 paper where reported using MSA software using gradients in PIPs PES for \ce{CH4}.\cite{ NandiQuBowman2019})  Instead, we typically use the entire data set (energies and optionally gradients as well) for fitting and then test the fit using out-of-sample data that are relevant to the application of interest.  Our reason for doing this is that we do over-determined linear regression, where the number of linear coefficients is much less than the size of the data set we fit, i.e., factors of 3-10.  In this way overfitting is avoided.  However, for the MD17 aspirin dataset which although large compared to the size of the PIPs basis is highly correlated, we do report some train and test metrics.

\subsection{Training and Testing Metrics}
We divided the rMD17 data set in to a training set (60\%) and a testing set (40\%) ensuring that data were chosen at random from the full set and that the training and testing set listed the geometries in a random order. These fits are all done with the 3$^{rd}$ order basis of 322221111111111 symmetry with 49 978 coefficients.  For the training, as shown in Table \ref{table:TrainTest}, we chose three combinations of gradients and energies, using the lowest (random) entries in the order of the training set geometries.  Fits were then made to the basis of order 3 with 49 978 coefficients/PIPs described in the next section.  The results of the fits are shown in the Train1, Train2, and Train 3 columns of the table. In each case, the test set consisted of 40 000 energies and gradient sets. The MAEs for both training and testing are based on fits of the weighted data described above, but there is no weighting involved in calculating the MAEs. In general the MAEs for the training and corresponding testing are close, even for the smallest training set consisting of 30 000 energies and 2000 gradients. Interestingly, the testing metrics improve only slightly as the number of gradients and energies in the training set is increased. 


The results of Test3 are not as good as the corresponding column (PES3) of Table \ref{table:PES}, which includes only the fit to the lowest-energy 4000 gradient sets and lowest 60 000 energies.

\begin{table}[htbp!]
    \centering
    \begin{tabular}{l c c c c c c}
    \toprule
    PES    &   Train1  &  Train2   &  Train3   & Test1  & Test2  & Test3  \\
    \midrule
   Number of Grads & 2000    &   3000   &  4000    &40 000    & 40 000   &  40 000     \\
   Number of Energies &  30 000  & 45 000    &  60 000   &  40 000   & 40 000    & 40,000     \\
   eRMSE (cm$^{-1}$) &  28    & 32  & 33  &  51   & 43  &    41    \\
   fRMSE (cm$^{-1}$/bohr) &  83     & 96  & 102   &  189   &  158 &   147   \\
   eWRMSE (cm$^{-1}$)&   27    & 31  & 32  &  48   &  42 &   39    \\
   fWRMSE\ (cm$^{-1}$/bohr) &  83    & 96  &  102  & 189    & 158  &   147    \\
   eMAE (cm$^{-1}$) &   21   & 24  & 25  & 36   &  32 &    30   \\
   fMAE (cm$^{-1}$/bohr)&  62     & 70  & 74  &  132   & 110  &    102   \\
   R$^2$(E)     &  0.999833     & 0.999785  & 0.999767    & 0.999440    & 0.999597  &  0.999644     \\
   R$^2$(G)     &  0.999765    & 0.999686  & 0.999643  & 0.999789    & 0.999152  &    0.999628   \\
    \bottomrule
    \end{tabular}
    \caption{Training and Testing for PIPs Potentials}
    \label{table:TrainTest}
\end{table}

\newpage
\subsection{The final PESs}
Having established robust testing and training using the larger PIP basis, Table \ref{table:PES} provides information on the potential energy surfaces, denoted PES1, PES2, PES3, and PES4, we constructed for further consideration.  MSA1 and MSA2 are the original outputs for $2^{nd}$- and $3^{rd}$-order polynomials; these were already described but they are included in the table for completeness.  No fits were done with these bases.  We then fit the smaller order basis to the lowest 80 000 energies and the 10 000 sets of gradients corresponding to the lowest energies to give the coefficients for PES1.  The rms error in energies was 123 cm$^{-1}$.  With more coefficients in the $3^{rd}$-order set of PIPS, we were able to reduce the rms error in energies to 23 - 27 cm$^{-1}$, with comparably good rms errors in the gradients.  Data on the number of gradients and energies for each of the three fits, PES2, PES3, and PES4 are provided in the table. In order to focus on the low-lying stationary points, in all three cases, we again used the geometries of the lowest energies and the gradients corresponding to the lowest energies.  Correlation plots for these surfaces are given in the Supporting Information (SI), and the $R^2$ correlation coefficients are shown in the table. In addition, all four PESs were provided with routines to perform analytic derivatives (fast) and reverse derivatives\cite{Houston2022PIPs} (very fast), as well as a routine to  calculate the Hessian using a combination of reverse and numerical derivatives. We note that the upper number of gradients and energies used in the fit is limited by the amount memory available to our computer nodes.  The minimum number of gradients and energies is set in order to avoid over-fitting: the data size from gradient components and energies needs to be at least 3-4 times the number of coefficients, or about 200 000 for the $3^{rd}$-order fits. 

\begin{table}[htbp!]
    \centering
    \begin{tabular}{l c c c c c c}
    \toprule
    PES:    &   MSA1  &   MSA2   &   PES1   & PES2  & PES3  & PES4  \\
    \midrule
   Polynomial order &  2  & 3 &  2-4  & 3 & 3 & 3  \\
   Num. coefficients     &  6766     &  297 952  & 25 001     & 49978 &  49978  &  49978  \\
   Number of G &  -   &   -   &  10 000   & 3500   &  4000   &  5000        \\
   Number of E &   -  &   -   &  80 000   & 60 000   & 60 000    & 80 000    \\   
   eRMSE(cm$^{-1}$)               &    -   &  -  & 123    &   23    &   23  &  27    \\
   fRMSE(cm$^{-1}$/bohr)          &    -   &  -  & 255    &   47    &   47  &  53    \\
   eWRMSE(cm$^{-1}$)              &    -   &  -  & 119    &   22    &   22  &  27    \\
   fWRMSE(cm$^{-1}$/bohr)         &    -   &  -  & 253    &   46    &   46  &  52    \\
   eMAE (cm$^{-1}$)               &    -   &  -  &  94    &   17    &   17  &  21    \\
   fMAE (cm$^{-1}$/bohr)          &    -   &  -  & 177    &   35    &   35  &  39    \\
   R$^2$(E)     &    -   &  - & 0.993577    &  0.999672   & 0.999672  &  0.999682    \\
   R$^2$(G)     &    -   &  - & 0.996879  &  0.999884   & 0.999884  & 0.999856       \\
    \bottomrule
    \end{tabular}
    \caption{Details of the PIPs potential energy surfaces and mean absolute errors for energies and gradients}
    \label{table:PES}
\end{table}

\subsection{Performance Metrics with Other Machine-Learned PESs}
\subsection{Fit Precision Comparisons}
Kov{\'a}cs \textit{et al}.\cite{Kovacs2021} considered fourteen ML methods for the MD17 aspirin dataset.\cite{Christensen2020,Schutt2021,gasteiger2022directional,NEURIPS2019,
Schnet,PhysNet,DTNN,GMdNN}   We refer the interested to reader to Table 1 and 2 of that paper for these results.  Here we focus on several of the most precise methods, including ACE, which was shown to be the most precise and efficient.

Data on the precision of the energy and forces for PES4 are provided in Table \ref{tab:PESMAE}, with comparisons to selected ML PESs trained on MD17 datasets for aspirin.  While comparisons are somewhat difficult due to different procedures, PES4 appears to outperform the others with respect to precision with the exception of most recent results using ``Allegro".\cite{Musaelian2023}. This is an atom-centered NN approach as is PaiNN.  ACE is also atom-centered but is a linear regression method.  More details of these methods are given below.  As seen, there are differences in the MAEs; however, arguably all these PESs provide very good precision and the factors of 2-3 among them are not very significant in our view.  However, as we shown below there are major differences in the computational efficiency.
 \begin{table}[htbp!]
    \centering
    \begin{tabular}{l l l }
    \toprule
    Property:    &   MAE  & MAE   \\
                &   Energy         &   Forces      \\
                &   cm$^{-1}$         &   cm$^{-1}$/bohr      \\
    PES        & & \\
    \midrule
        PES1$^1$ (this work) & 94  & 77 \\ 
        PES4$^1$ (this work)    & 21  & 39 \\ 
        ACE$^3$    & 49  & 76 \\
        PaiNN$^{4}$  & 56    & 76   \\
        Allegro$^5$  & 18  & 31 \\
    \bottomrule
    \end{tabular}
    \begin{tablenotes}
    \item[$^1$] $^1$Basis of 25 001 PIPs
    \item[$^2$] $^2$Basis of 49 978 PIPs
    \item[$^3$] $^3$Reference \citenum{Kovacs2021} using rMD17, training on 1000 geometries
    \item[$^4$] $^4$Reference \citenum{Schutt2021} using MD17, training on 1000 geometries
        \item[$5$] $^5$Reference \citenum{Musaelian2023} using rMD17

    \end{tablenotes}
    \caption{Accuracy of Energies and Forces}
    \label{tab:PESMAE}
\end{table}

 A final test of the precision of PES4 is provided by an examination of properties of the stationary points on the PES4 surface, with  comparisons to the PBE/def2-SVP Molpro results and those from the ACE surface.  The relative energy of the TS1 barrier to methyl rotation is 295 \cm on PES4 and 234 \cm from the PBE/def2-SVP benchmark.  Given the sparsity of points in the energy range, this seems acceptable.  The mean absolute error (MAE) for all of the 54 vibrational frequencies at the GM, is 5.0 \cm for PES4, and 2.2 and 2.4 \cm for ACE when fitting using forces or energies and forces, respectively.  All surfaces have reasonable agreement with the benchmark used in ref.\citenum{Kovacs2021}, although ACE outperforms PES4 on this metric. We also determined the vibrational frequencies of TS1, the transition state for methyl rotation (described below) on the PES4 surface and found a similar MAE, 5.0 \cm, with the benchmark we calculated using PBE/def2-SVP and Molpro.\cite{MOLPRO_brief}

\subsection{Timing Comparisons Energies and Forces}
Separate timing measurements for 100,000 gradient sets and energies were performed using PES4 and yielded 39.51 s and 12.0 s, respectively, using the reverse differentiation.  Note that the ratio of gradient to energy time is about 3.3, consistent with Fig. 6 of ref. \citenum{Houston2022PIPs}, which indicated that for reverse differentiation this time ratio is independent of the number of atoms in the molecule.  These numbers correspond to about 0.4 ms per gradient set, and 0.12 ms per energy.  Given that there are 21 atoms in aspirin, the gradient time is equivalent to 0.019 ms per atom. 

\begin{figure}[htbp!]
\begin{center}
\includegraphics[width=.7\textwidth]{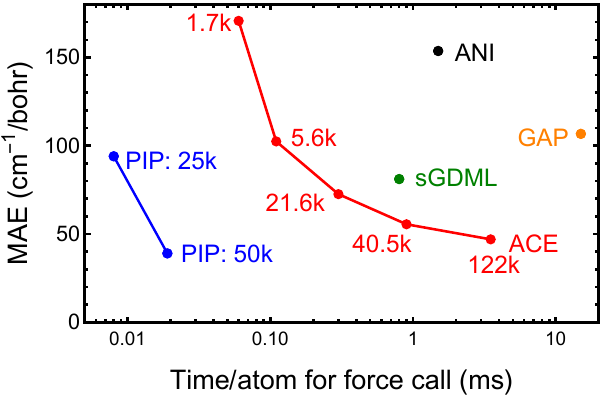}
\end{center}
\caption{A semi-log plot comparing the MAE for calculated vs. ab initio forces as a function of the time per atom for determining the force or gradient. Results of this work (on aspirin) are shown in blue; others (on azobenzene) are taken from Fig. 2 of ref. \citenum{Kovacs2021}.}
\label{fig:timing}
\end{figure}

One of the fastest and most accurate of other ML potential energy surfaces is ACE, so it is of interest to make a comparison to a recent result.\cite{Kovacs2021}  Although these authors did not report timing data for aspirin, they did so for azobenzene, with 24 atoms and one of the basis sizes (40.5k) was similar to that of the largest basis used for aspirin (about 50k). Further, the timings are per atom and this means the comparisons shown are as consistent as possible.  To get a fit with MAE $\approx$ 15 meV/\AA \hspace{1em}(64 \cm/bohr), they used 40,500 basis functions and required 1 ms per atom to get the forces (negative gradients).  Their MAE of 64 \cm/bohr is larger than ours for PES4 (39 \cm/bohr) for a similar number of basis functions (40,500 vs 49,978).  Yet for PES4, our time per atom for the gradients (0.019 ms) is about 50 times lower than theirs (1 ms).  Their computer, a 2.3 GHz Intel Xeon Gold 5218, has about the same single-core performance as the Xeon Gold 6250 used in our measurements. Because PES2, PES3, and Train1, Train2, Train3 all use the same basis as PES4, they will all have the same timing.  Our data on aspirin and data from others on azobenzene, taken from Fig 2 of ref.  \citenum{Kovacs2021}, are shown in Fig. \ref{fig:timing}.  At least for this comparison, it appears that the PIP method is both more accurate and much faster than the others.  

\section{Beyond MAEs}
\subsection{Potential and Hindered Rotational Frequencies of the Methyl Rotor}
 Once a satisfactory PES has been fit, we are then in a position to evaluate important minima and transition states.  The global minimum (GM) is usually fairly easy to locate by following an gradient path until the gradients become negligible. Transition states are somewhat harder to locate.  With a good guess as to the transition state geometry, Molpro or Gaussian can often locate transition states based on the fact that the gradients will all be zero and that at least one vibrational frequency will be imaginary.  Similar techniques can be used on the PES.  In either case, development of a good starting guess often requires examination of several geometries related the the motion across the transition state.  For the examination of the methyl rotation, we found coordinates that were rotated from those in the global minimum structure by a) transforming the origin of the geometries to be at the center of mass of the three hydrogen atoms, b) solving for the Euler angles that rotated the aspirin about this origin such that the three hydrogens were in the $y-z$ plane, and then c) calculating positions of the three hydrogen atoms as a function of the rotational angle $\phi$ in this plane, as measured from the positions of the atoms in the global minimum structure.  We then calculated the energies of these points on the PES and took one of the three geometries with the maximum-energy structure as an estimate of the transition state structure. This enabled us to find the actual structure both in a benchmark Molpro calculation and on the PES.  The actual structure was the one in which the angle $\phi$ was that expected, $60^{\circ}$ from the GM, but in which the remainder of the molecule was ``relaxed'' into its minimum energy geometry.

 We determined the intrinsic reaction coordinate
 using a  method in which a trajectory is started at the transition state with a small amount of kinetic energy.  The kinetic energy is diminished by a factor of 0.98 at each time step; additionally, a ``brake'' factor of 0.5 is applied on the next step if the kinetic energy exceeds 20 \cm.    
 This guides the trajectory along the path of steepest descent. The trajectory eventually moves to the nearest local or global minimum. About 60,000 steps at 3 a.u. in time per step were performed.

\begin{figure}[htbp!]
\begin{center}
\includegraphics[width=.7\textwidth]{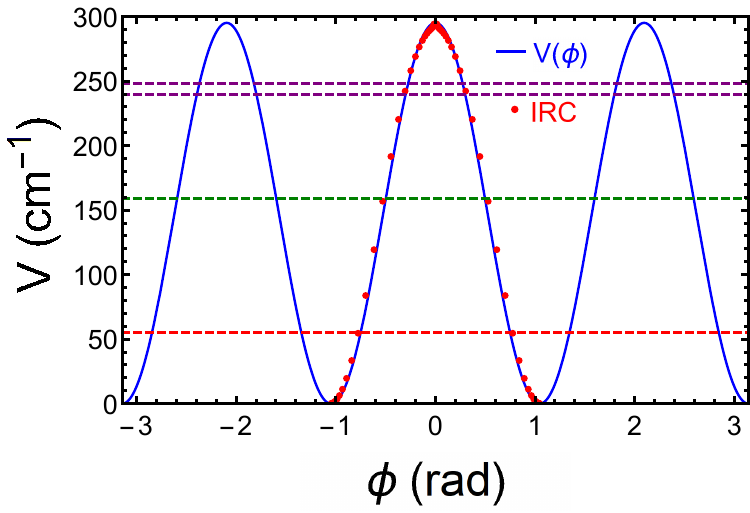}
\end{center}
\caption{The torsional potential for methyl rotation in aspirin, showing the intrinsic reaction coordinate (red) and $V(\phi)=295 (.5 \text{Cos}(3\phi))+.5)$ is an excellent approximation to the potential. }
\label{fig:CH3rotation}
\end{figure}

Figure \ref{fig:CH3rotation} shows a one-dimensional cut of PES4 along the coordinate for methyl rotation. The red points show the minimum energy path, calculated by the trajectory method described above. The barrier to internal rotation is about 295 cm$^{-1}$, and the potential is very accurately fit by the function $V(\phi)=295 (.5 \text{Cos}(3\phi))+.5)$. The harmonic frequency corresponding to this mode at the GM is 128 cm$^{-1}$, whereas the the lowest-energy transition frequency determined by using a 1d DVR calculation\cite{ColbertMiller1992} is 103 \cm.

\subsubsection{H-atom vibrational energies from the formic acid group}
\begin{figure}[htbp!]
\begin{center}
\includegraphics[width=.7\textwidth]{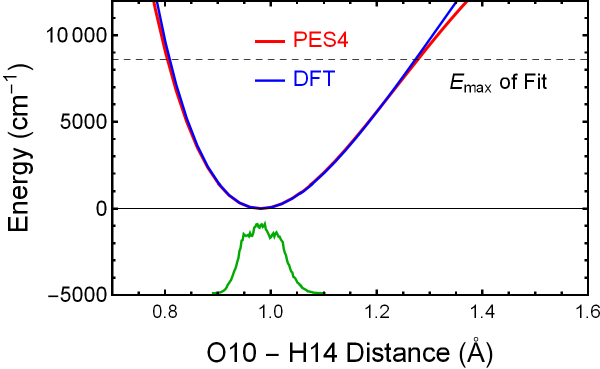}
\end{center}
\caption{Displacement of H14 along the O10-H14 bond (see Fig. \ref{fig:ballstick} for identifications). The PES4 results are shown in red, while the DFT benchmark results are in blue. The maximum total energy used in the fit is shown as the black dashed line. The distribution of the O10-H14 distances in the data set is shown by the green curve.  }
\label{fig:hatomdissoc}
\end{figure}

The potential for removing the H atom on the formic acid group along the O10-H14 bond (see Fig. \ref{fig:ballstick} for identities) is shown as the red curve in Fig. \ref{fig:hatomdissoc} along with the results of a DFT PBE/def2-SVP calculation of the energies at the same geometries.  Also shown are the O10-H14 distance distribution of the data set (green), and the maximum energy used in the fitting (dashed, black). This is an unrelaxed potential energy cut, but it is unlikely that the minimum energy path is very much different.

PES4 extrapolates well on both sides of the OH distance for the global minimum even though there is very little corresponding data for these O-H distances in the data set.  However, as the energy approaches the maximum total energy used for the fit, there are deviations, slight ones on the lower side but larger ones on the upper side.  The reason for the deviations is that there are no data points at energies that would constrain the polynomial fit. Without such data the fitting will let the polynomial take whatever shape best fits the existing data points.  In this case, the PES cut turns over near 1.7 \AA~  and heads toward a large negative number, creating what we call a ``hole.''  If we wanted to describe the H atom dissociation at longer distances than those in the diagram,  this hole could readily be filled by adding points to the dataset at higher energies along this cut. The rMD17 data set contains very few points above the maximum energy of the fit, and none with the energy in this range of the O10-H14 distance.  

We should emphasize that this motion is not close to a normal mode of the global minimum configuration.  Rather, it is a local mode, a well-established model that was applied to an analysis of overtones of OH-stretches in \ce{H2O}\cite{child79}. A 1-d DVR calculation of vibrational energies yields a zero point for this motion of 1830 \cm, and a fundamental excitation energy of 3742 \cm.

\newpage
\subsubsection{Rotation of the formic acid group}
\begin{figure}[htbp!]
\begin{center}
\includegraphics[width=.7\textwidth]{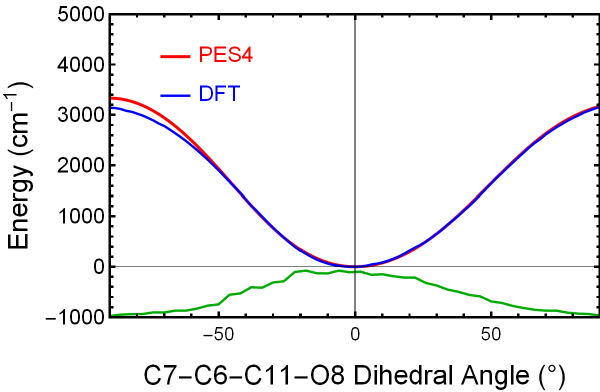}
\end{center}
\caption{Potential energy as a function of the dihedral angle C7-C6-C11-O8 (see Fig. \ref{fig:ballstick} for identifications). The potential energy for PES4 is shown in red, the DFT results are shown in blue, and the distribution of this dihedral angle in the data set is shown by the green curve.}
\label{fig:formicrot}
\end{figure}

Another property of interest is the rotation of the formic acid group around the C6-C11 bond.  Figure \ref{fig:formicrot} shows the potential energy in red and the DFT energy in blue for this rotation, where the abscissa is the  C7-C6-C11-O8 dihedral angle. The green curve gives the distribution of this angle in the data set.  As can be seen from the figure, PES4 and the DFT energies agree very well except for a slight deviation at -60 to -90 degrees where structures in the data set are rather sparse.  Vibrations perpendicular to this path have not been optimized, so this is not a minimum energy path. A 1-d DVR calculation gives an approximate zero-point energy of 36.5 \cm and a vibrational quantum of 72.9 \cm.   

\newpage
\subsubsection{Rotation of the acetyl group}

\begin{figure}[htbp!]
\begin{center}
\includegraphics[width=.8\textwidth]{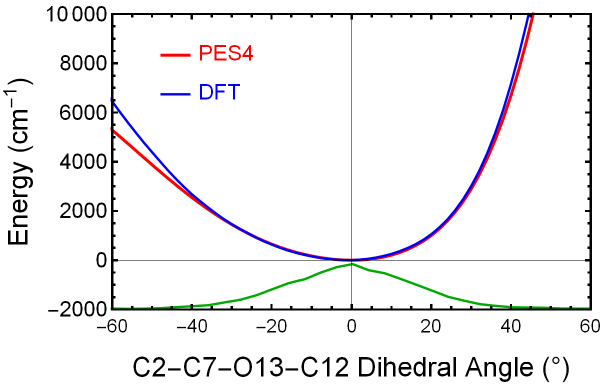}
\end{center}
\caption{Potential energy as a function of the dihedral angle C2-C7-O13-C12 (see Fig. \ref{fig:ballstick} for identifications). The potential energy for PES4 is shown in red, the DFT results are shown in blue, and the distribution of this dihedral angle in the data set is shown by the green curve.}
\label{fig:CH3COrot}
\end{figure}

The potential energy as a function of the dihedral angle C2-C7-O13-C12 is shown in Fig \ref{fig:CH3COrot}.  This gives the rotational potential of the CH3CO group around the C7-O13 axis. As in previous cases, the energies from PES4 and those from DFT are in good agreement until the angles reach geometries that are not well represented in the data set, whose distribution is shown in the green curve.   

\newpage
\section{Discussion}
The most notable result of this study is the speed of evaluation of the PIPs PES for aspirin compared to all other ML considered here.  The fit precision for energies and forces are lower than all ML methods shown in Fig. \ref{fig:timing}.  The MAE differences shown in that figure for forces are within a factor of five; however, the differences in timing compared to the PIPs PES are at least an order of magnitude bigger, i.e., factor of 50 or more.  So it is the large difference in the evaluation speed that deserves some discussion. Before doing that, we note that timing assessments like the ones shown that figure were reported earlier by us for ethanol, where again factors of 12-117 were noted with respect to ANI and 12.3 with respect to ACE.\cite{Houston2022PIPs}

First consider the atom-based methods, ANI, GAP and ACE. These methods all scale with the number of atoms, although with different pre-factors; however, the the timing differences are within a factor of 5 or so, i.e., the same range as the differences in the MAE.   ANI, the least precise of the methods included in Figure \ref{fig:timing}, is an elaborated version of the original and seminal atom-centered NN approach of Behler and Parinello.\cite{BP,behl1,behlpccp}.  GAP is based on Gaussian Process Regression.  Both scale non-linearly with the parameters that are optimized to minimize the loss. This is also true for sGDML, which is not an atom-centered method, but which shares in common with GAP the need both to evaluate a matrix inverse in training and then also perform a matrix multiplication.    ACE is a liner regression method, like PIPs, where the prefactor scales with the size of the basis.  However, the large difference between ACE and PIPs is that PIPs is not atom-centered.  A single basis is used to represent the PES. (It should be noted that this comment also applies to the highly successful PIP-NN\cite{guo20} and PIP/FI-NN\cite{nsr2023} methods and to the automated PIP software ``Robosurfer"\cite{czako2021}.) So for the same size basis the cpu effort for ACE scales roughly like N times the cpu time for PIPs. In the present case of aspirin the factor is 21.  This is indeed seen for the case of a PIP basis of 50 000 compared to the ACE basis of 40 500.  However, to achieve an MAE almost as small as PIPs requires a basis twice the size of the PIPs basis and so the timing difference is close to 50.  

Beyond timing and MAEs the present PES extrapolates accurately and sufficiently beyond the range of data to enable a an accurate (albeit approximate) calculation of the OH-stretch fundamental, compared to using direct DFT energies for that cut.  Similarly the PES compares very well with direct DFT energies  beyond the range of data for two internal rotations.  

It seems appropriate to ask whether it will be possible with PIPs to calculate PESs for much larger molecules than aspirin.  Many suppose that since the number of internuclear distances expands as $N(N-1)/2$ where $N$ is the number of atoms, then the basis set will grow to be completely unwieldy.  What this superficial analysis fails to consider is that many Morse (or reciprocal) transforms of the internuclear distances will be very small when the molecule gets large and can therefore can be neglected by using either fragmentation of the basis into parts\cite{NandiBowman2019} or by pruning the basis set to remove them.\cite{Houston2023PIPsoftware}  It is instructive to examine the scaling of internuclear distance with number of atoms.  Consider the volume, $V$, of a globular molecule; it increases as $N^3$ so that $N \sim V^{1/3}$.  The radius also varies as $V^{1/3}$, and it follows that $R \sim N$; i.e, $R$ varies linearly with $N$. Now suppose that we can neglect any transformed internuclear distance that is larger than a fixed distance $d_{max}$.  As $N$ grows, so does $R$, but the number of internuclear distances within a radius of $d_{max}$ stays the same.  We may have to move the center of the sphere of radius $d_{max}$ to cover the complete molecule, as we essentially do using fragmentation, but the basis size will remain the same, though applied with different groups of internuclear distances.  We thus conclude that is is not likely to be the size of the PIP basis that makes things difficult for application to larger molecules.

It is actually the size of the required dataset that causes the problems, particularly for molecules as opposed to structured systems such as solids, surfaces, self-assembled structures, etc., where one might take advantage of translational symmetry.  As a molecule grows in size, more phase space for the atoms becomes available, and the dataset must cover a larger number of configurations. Because accurate electronic structure calculations are already very expensive for molecules of the size of Aspirin, progress with the potential energy surfaces of larger molecules will stall unless major advances in these calculations can be achieved.  This is an interesting reversal in the field.  For many years the bottleneck in creating full-dimensional PESs was the lack of methodology to go beyond tetraatomic molecules, while electronic structure methodology, notably DFT, could generate lots of data for molecules 2-4 times that size.  Now that the capability to develop MLPs for  molecules with more than 20 atoms is becoming routine it is the electronic structure step that has become the rate-limiting; the case of quantum computing thus becomes even more compelling for Computational Chemistry. 

\section{Summary and Conclusions}
Several PIP PESs for 21-atom aspirin trained on the rMD17 data set were reported.  This extends the PIP approach to the largest molecule to date and adds the PIPs approach to the large number of ML methods to fit this dataset. The PES outperforms all other ML potentials reported for the molecule and dataset, with respect to fitting precision and more dramatically with respect to speed of evaluation.  Specifically the PIPs potential runs roughly 50 times faster than ACE and with a smaller MAE for energies and forces.  The former and more important difference was traced to a fundamental difference in the PIPs and ACE (``atomic centered expansion") architecture.  In the PIPs approach a single global basis of PIPs is used to represent the full potential.  This is in contrast to ACE (and all ``atom-centered'' approaches), which express the potential as a sum of ``atomic" energies.  In the ACE implementation each of these is expressed in terms of a large number of basis functions.  The number used to achieve the ultimate MAE is 122 000, which is more than twice the size of the single PIPs basis.  Then multiplying that factor by the number atom atoms (21) gives the factor or roughly 50 reported as the speed ratio.  

Several properties of the PES were reported for the first time, namely the methyl torsional potential,  and the 1d anharmonic vibrational energies of the local mode OH stretch. This and other large amplitude motions are accurately described by the PES in regions beyond the dataset in comparison to direct DFT calculations.  

The major conclusion of this work is the speed advantage of the PIPs approach compared to the atom-centered approaches for 21-atom aspirin.  This extends an earlier and identical conclusion for the 9-atom ethanol and leads us to conclude that at the present time the PIPs approach is the fastest, high-precision method to represent molecules, now with up to 21 atoms.  Looking ahead there are still challenges for the PIPs approach to molecules with substantially more than 20 atoms and the software to obtain the PIPs basis.  The fragmentation approach, introduced and tested on several molecules including $N$-methyl acetamide\cite{NandiQuBowman2019}, AcAc\cite{QuAcAc2021} and tropolone\cite{Nandi2023JACS}, appears to be a very viable approach for large molecules.  However, this remains for future work.
 
It is now possible to develop semi-global or global Machine-Learned Potentials for molecules with more than 20 atoms. Unlike direct dynamics approaches, MLPs permit long-time dynamical and statistical mechanical simulations of a molecular motion.  The ``rate-limiting step" now appears to be the speed of electronic structure methods to generate the tens of thousands of energies (and gradients perhaps) is clear.  This is testament to the remarkable progress made in developing MLPs over the past 15 years. 

\section{Data Availability}
The PESs used in this study are available by contacting the authors.

\section{Supporting Information}
\begin{itemize}
\item Structural Information for the GM and TS1
\item Harmonic vibrational Frequencies for the GM and TS1
\item Correlation Plots for PES2, PES3, and PES4

\end{itemize}
\begin{acknowledgement}

%
 The authors thank David Kovács and Gábor Csányi for useful discussions. JMB thanks the Army Research Office, DURIP grant (W911NF-14-1-0471), for funding a computer cluster where most of the calculations were performed. JMB and PP acknowledge current support from NASA grant (80NSSC22K1167).

\end{acknowledgement}


\bibliography{refs}



\newpage
\section{Supporting Information}

\section{Structural Infomation (coordinates in \AA)}
\subsection{GM from Molpro calculations}   
    21 \\
 RKS-SCF000/DEF2-SVP  ENERGY=-647.49258073 \\
 C          1.6325220252       -1.4206418605       -0.7116536172 \\
 C          0.8610038938        1.0628815086       -1.7422482356 \\
 C          2.4167875455       -0.8026233507       -1.6911352236 \\
 C          2.0250521475        0.4418052894       -2.2119447131 \\
 C         -2.6415527769        1.9294250128        1.1877285005 \\
 C          0.4483684828       -0.8186228399       -0.2275050207 \\
 C          0.0750216504        0.4449468999       -0.7574336563 \\
 O         -1.4839327118       -1.1557719182        1.1992051464 \\
 O         -0.2774188184        1.6277401749        1.6802321074 \\
 O          0.1416437076       -2.7132776796        1.1633773322 \\
 C         -0.4048978786       -1.5323783011        0.7667937290 \\
 C         -1.2119166772        1.5529687278        0.9212447348 \\
 O         -1.0876215807        1.0855513971       -0.3869674614 \\
 H         -0.5033981755       -3.0836550032        1.8033681192 \\
 H          1.9226767656       -2.3980125525       -0.3030291473 \\
 H          0.5352517172        2.0402316532       -2.1263312466 \\
 H          3.3329521253       -1.2932014041       -2.0518145322 \\
 H          2.6325806473        0.9365586583       -2.9851413416 \\
 H         -2.6985753123        2.5791597925        2.0781191942 \\
 H         -3.1023594152        2.4169613260        0.3082766959 \\
 H         -3.1939366218        0.9872526993        1.3816364660 \\

 \subsection{GM from PES4}

 21 \\
  -647.49337286      \\
C    -2.568497702    0.762223566    3.751404474 \\
C    -3.348848282    3.243119570    2.724444566 \\
C    -1.784344979    1.386447440    2.776712330 \\
C    -2.180123809    2.629948108    2.258375295 \\
C    -6.849828102    4.112113191    5.643526165 \\
C    -3.756847680    1.357026545    4.233278764 \\
C    -4.134650253    2.619459404    3.705063022 \\
O    -5.687842786    1.011623331    5.658038481 \\
O    -4.481493520    3.829816304    6.126881656 \\
O    -4.056247968   -0.539238272    5.623240239 \\
C    -4.607530903    0.639288956    5.226861056 \\
C    -5.420001785    3.739043306    5.374770623 \\
O    -5.302000157    3.253044307    4.073182763 \\
H    -4.700436833   -0.911191744    6.262955248 \\
H    -2.275235503   -0.214186313    4.159090096 \\
H    -3.679365565    4.218140175    2.339945304 \\
H    -0.864909451    0.902006374    2.417223448 \\
H    -1.572532287    3.129279586    1.488822100 \\
H    -6.906491042    4.752817130    6.539701166 \\
H    -7.305818879    4.610180325    4.767919704 \\
H    -7.403735925    3.169398799    5.825267134 \\ 
\section{Evaluation of PES Fits}

\subsection{TS1 from Molpro calculations}

   21 \\
 RKS-SCF000/DEF2-SVP  ENERGY=-647.49151450 \\
 C          5.5565632284       -2.3106069106       -0.9149520201 \\
 C          4.4831546390       -0.0423707054       -2.1535035263 \\
 C          6.3681915247       -1.5168804611       -1.7323048914 \\
 C          5.8264496328       -0.3819588366       -2.3583614070 \\
 C          0.3570562694       -0.0037579301        0.0009425391 \\
 C          4.1957236735       -1.9944132134       -0.6992772316 \\
 C          3.6699382458       -0.8364341457       -1.3309023815 \\
 O          2.1355635686       -2.7670197929        0.3224852880 \\
 O          2.5824835664        0.1700025509        0.9573473632 \\
 O          4.0310844115       -3.9393122941        0.6454552651 \\
 C          3.3340613389       -2.8950499792        0.1213641548 \\
 C          1.8587714971       -0.0967377278        0.0296366438 \\
 O          2.3429059490       -0.4815342626       -1.2193995332 \\
 H          3.3664561657       -4.4515666391        1.1541125936 \\
 H          5.9651920910       -3.2060263928       -0.4274093939 \\
 H          4.0354584383        0.8446376438       -2.6241321729 \\
 H          7.4240902630       -1.7851737912       -1.8845894640 \\
 H          6.4549274959        0.2480718162       -3.0062912719 \\
 H          0.0581613216        1.0626247251       -0.0600089104 \\
 H         -0.0709340675       -0.5535189731       -0.8549548895 \\
 H         -0.0329158649       -0.4091343491        0.9528489587 \\
 
\subsection{TS1 from PES4}

 21 \\
  -647.4920510398      \\
C     2.226160434    3.471954713    7.805352950 \\
C     1.131548147    5.733808185    6.576900611 \\
C     3.029771250    4.276671846    6.992022358 \\
C     2.477735835    5.409307769    6.372208100 \\
C    -2.989800078    5.784872443    8.709761077 \\
C     0.862935728    3.774569669    8.023368249 \\
C     0.326008366    4.929080547    7.396057223 \\
O    -1.186877621    2.984951082    9.048175852 \\
O    -0.768057636    5.951663432    9.673683232 \\
O     0.718397779    1.826935356    9.366463076 \\
C     0.011292073    2.865727024    8.844987824 \\
C    -1.489206866    5.679682639    8.745646633 \\
O    -1.005110136    5.271188243    7.505452742 \\
H     0.058296562    1.309335210    9.875094561 \\
H     2.643268151    2.578787346    8.288977833 \\
H     0.674728431    6.616177739    6.107603518 \\
H     4.087545481    4.019125074    6.837303694 \\
H     3.100012532    6.047583124    5.727354976 \\
H    -3.278034466    6.852926019    8.640748364 \\
H    -3.418284429    5.232761659    7.856390710 \\
H    -3.385863087    5.389516671    9.662569894 \\

\section{Harmonic Frequencies (in cm$^{-1}$)}
\subsection{DFT and PES4 Results for the GM}

\begin{table}[htbp!]
\centering
\caption{Harmonic Frequencies for the GM from DFT and PES4}
\label{tab:gmvibs}
\begin{threeparttable}
\begin{tabular*}{0.8\columnwidth}{@{\extracolsep{\fill}} c c c c c c }
\hline
\hline\noalign{\smallskip}
  Mode &  DFT & PES & Mode & DFT & PES  \\  
  \hline\noalign{\smallskip}
   1 & 33.6  &  34.7 & 29 & 984.2    &  987.9 \\
   2 & 83.1  &  85.1 & 30 & 998.3    &  1012.9 \\
   3 & 84.9  &  90.4 & 31 & 1037.5  &  1039.8\\
   4 & 100.9 & 100.8  & 32 & 1060.7  &  1066.0\\
   5 & 115.8 & 128.2   & 33 & 1118.0 &  1123.3 \\
   6 & 137.5 & 138.0   & 34 & 1133.6  &  1139.3\\
   7 & 224.6  & 226.3  & 35 & 1160.0  &  1168.0\\
   8 & 261.7  & 262.9  & 36 & 1168.3  &  1173.5\\
   9 & 302.6  & 305.7  & 37 & 1222.9  &  1228.2\\
   10 & 352.8  &  354.5 & 38 & 1244.6  &  1251.7\\
   11 & 413.1  & 414.2  & 39 & 1315.1 &  1332.5\\
   12 & 425.2  & 426.2  & 40 & 1341.5 &  1352.3\\
   13 & 501.5  & 502.8  & 41 & 1383.1 &  1385.4\\
   14 & 515.0  & 519.2  & 42 & 1388.6 &  1405.1\\
   15 & 542.4  & 545.1  & 43 & 1393.4  &  1410.6\\
   16 & 554.4  & 563.2  & 44 & 1436.4 &  1439.8\\
   17 & 598.6  & 594.6  & 45 & 1473.9 &  1479.5\\
   18 & 624.6  & 628.6  & 46 & 1588.7 &  1593.6\\
   19 & 655.9  & 656.9  & 47 & 1620.8  &  1623.7\\
   20 & 705.6  & 707.0  & 48 & 1767.2 &  1770.5\\
   21 & 726.1  & 727.7  & 49 & 1833.9 &  1837.3\\
   22 & 757.0  & 760.0  & 50 & 2995.6 &  3005.6\\
   23 & 791.2  & 793.1  & 51 & 3087.4 &  3097.6\\
   24 & 810.7  & 812.5  & 52 & 3119.6  &  3128.2\\
   25 & 872.7  & 877.2  & 53 & 3127.8 &  3135.9\\
   26 & 904.15  & 907.2  & 54 & 3134.4  &  3146.0\\
   27 & 953.1  & 961.7  & 55 & 3144.3&  3150.1\\
   28 & 970.3  & 978.6  & 56 & 3157.3 &  3163.6\\
    &  &   & 57 & 3638.3 &  3653.3\\
\hline\noalign{\smallskip}
& MAE = 5.55 cm$^{-1}$ &\\
\hline
\hline\noalign{\smallskip}
\end{tabular*}
\end{threeparttable}
\end{table}

\newpage
\subsection{DFT and PES4 Results for TS1}

\begin{table}[htbp!]
\centering
\caption{Harmonic Frequencies for the TS1 from DFT and PES4}
\label{tab:gmvibs}
\begin{threeparttable}
\begin{tabular*}{.8 \columnwidth}{@{\extracolsep{\fill}} c c c c c c }
\hline
\hline\noalign{\smallskip}
  Mode &  DFT & PES & Mode & DFT & PES  \\  
  \hline\noalign{\smallskip}
   1 & 108.0 i & 113.4 i & 29 & 988.0 & 984.5 \\
 2 & 35.7 & 39.1 & 30 & 1011.0 & 998.6 \\
 3 & 69.4 & 73.9 & 31 & 1038.8 & 1037.0 \\
 4 & 89.4 & 90.8 & 32 & 1066.9 & 1061.0 \\
 5 & 100.6 & 104.3 & 33 & 1121.6 & 1117.1 \\
 6 & 139.9 & 140.6 & 34 & 1134.7 & 1134.1 \\
 7 & 226.1 & 228.4 & 35 & 1166.3 & 1156.8 \\
 8 & 264.9 & 263.4 & 36 & 1172.7 & 1167.9 \\
 9 & 309.8 & 309.1 & 37 & 1225.5 & 1223.1 \\
 10 & 354.8 & 353.7 & 38 & 1251.2 & 1244.7 \\
 11 & 417.2 & 415.9 & 39 & 1333.8 & 1319.0 \\
 12 & 426.6 & 425.9 & 40 & 1349.8 & 1341.7 \\
 13 & 502.6 & 502.1 & 41 & 1385.1 & 1383.4 \\
 14 & 523.6 & 521.1 & 42 & 1401.4 & 1387.3 \\
 15 & 540.2 & 537.7 & 43 & 1412.9 & 1397.6 \\
 16 & 575.9 & 567.8 & 44 & 1440.9 & 1436.9 \\
 17 & 594.8 & 598.6 & 45 & 1480.5 & 1473.9 \\
 18 & 628.5 & 625.4 & 46 & 1593.9 & 1589.0 \\
 19 & 653.7 & 652.9 & 47 & 1624.6 & 1620.9 \\
 20 & 706.7 & 705.8 & 48 & 1769.8 & 1765.7 \\
 21 & 728.0 & 727.9 & 49 & 1834.2 & 1831.0 \\
 22 & 756.9 & 755.8 & 50 & 3008.0 & 2997.9 \\
 23 & 794.0 & 792.3 & 51 & 3104.5 & 3089.8 \\
 24 & 815.0 & 813.4 & 52 & 3131.3 & 3119.4 \\
 25 & 876.8 & 872.7 & 53 & 3134.6 & 3128.6 \\
 26 & 904.4 & 901.5 & 54 & 3145.3 & 3134.3 \\
 27 & 961.2 & 953.1 & 55 & 3151.2 & 3144.9 \\
 28 & 983.2 & 980.0 & 56 & 3162.6 & 3157.3 \\
    &  &   & 57 & 3638.2 &  3651.5\\
\hline\noalign{\smallskip}
& MAE = 4.96 cm$^{-1}$ &\\
\hline
\hline\noalign{\smallskip}
\end{tabular*}
\end{threeparttable}
\end{table}

\newpage
\section{Correlation plots for PES2, PES3, and PES4}
\subsection{Correlation plot for PES2}

\begin{figure}[htbp!]
\begin{center}
\includegraphics[width=.6\textwidth]{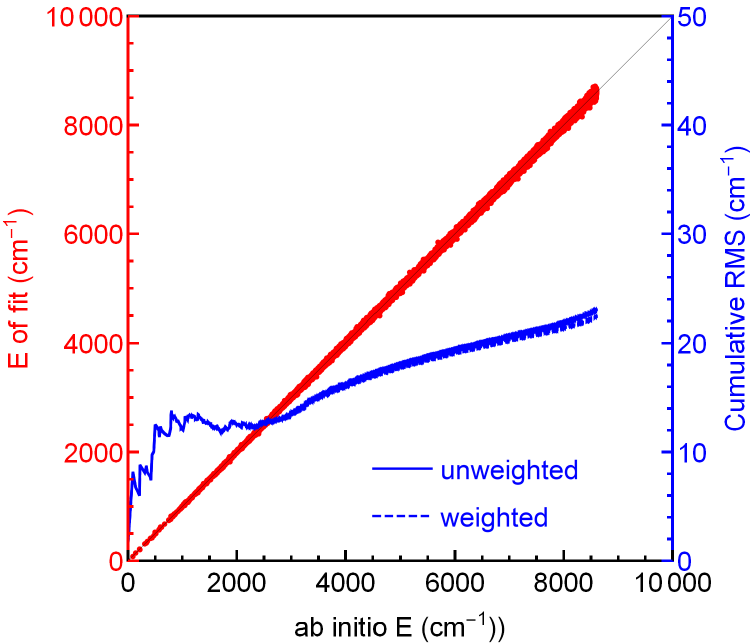}
\includegraphics[width=.6\textwidth]{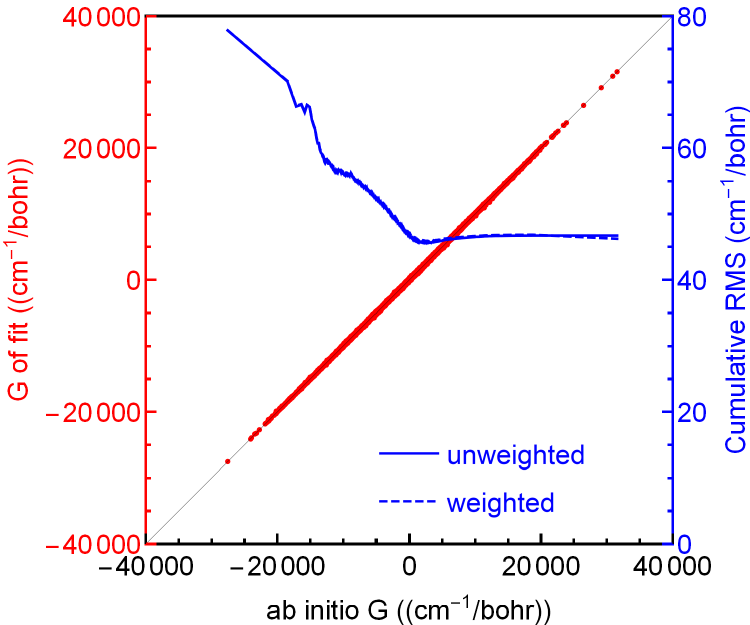}
\end{center}
\caption{ PES2}
\end{figure}

\newpage
\subsection{Correlation plot for PES3}

\begin{figure}[htbp!]
\begin{center}
\includegraphics[width=.6\textwidth]{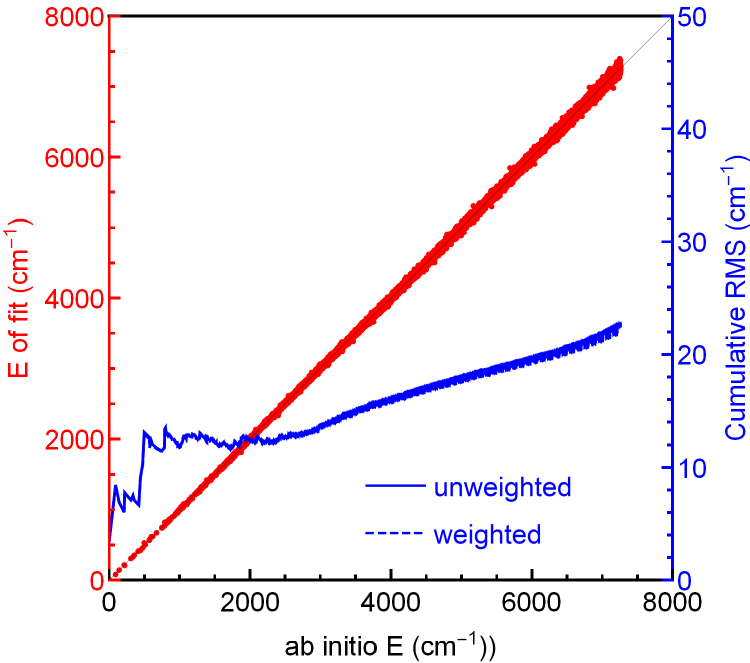}
\includegraphics[width=.6\textwidth]{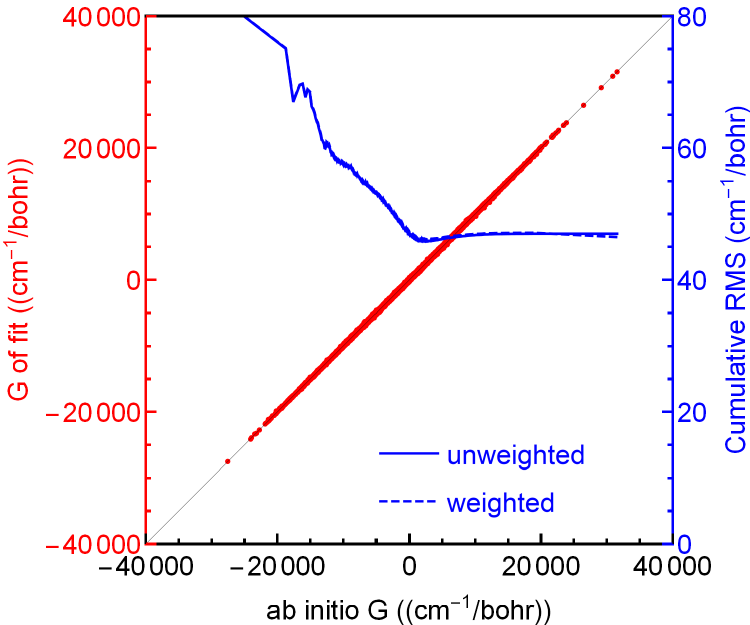}
\end{center}
\caption{ PES3}
\end{figure}

\newpage
\subsection{Correlation plot for PES4}

\begin{figure}[htbp!]
\begin{center}
\includegraphics[width=.6\textwidth]{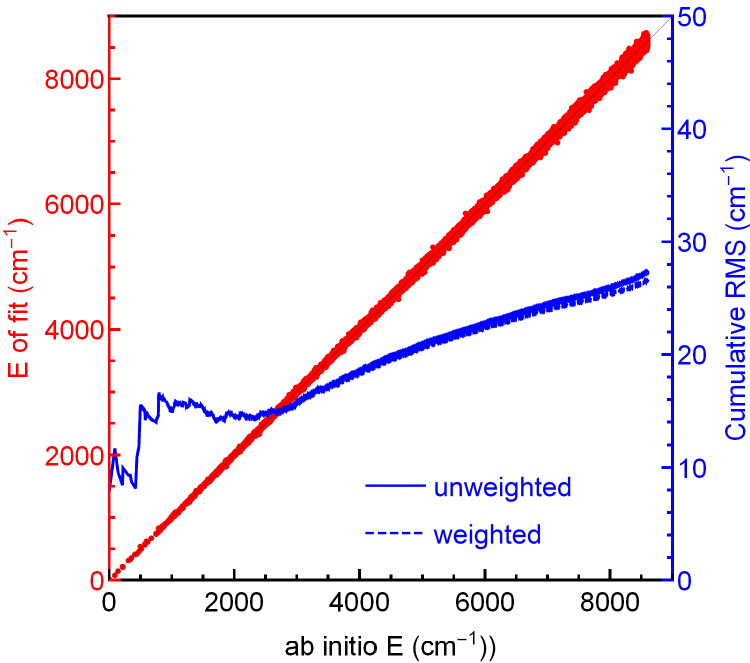}
\includegraphics[width=.6\textwidth]{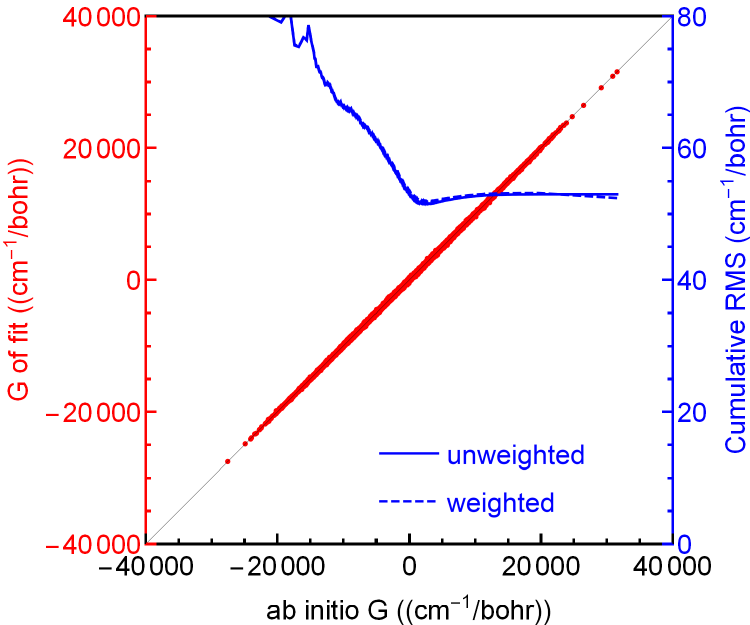}
\end{center}
\caption{ PES4}
\end{figure}






\end{document}